\documentclass[12pt]{article}
\input{epsf.tex}
\def\G{{\cal G}^{+++}}
\begin{document}
\thispagestyle{empty}
\setcounter{page}{0}
\renewcommand{\theequation}{\thesection.\arabic{equation}}

{\hfill{\tt hep-th/0405082}}

{\hfill{ULB-TH/04-12}}

\vspace{1.5cm}

\begin{center}  {\bf    $\G$-INVARIANT FORMULATION OF GRAVITY AND 
M-THEORIES:  EXACT INTERSECTING BRANE SOLUTIONS} 

\vspace{.5cm}

Fran\c cois Englert${}^{a,c}$ and Laurent Houart${}^{b,c}$\footnote{Research Associate
F.N.R.S.}

\footnotesize \vspace{.5 cm}

${}^a${\em Service de Physique Th\'eorique, Universit\'e Libre de Bruxelles,
\\ Campus Plaine, C.P.225\\Boulevard du Triomphe, B-1050 Bruxelles, Belgium}\\  {\tt
fenglert@ulb.ac.be}

\vspace{.2cm}

${}^b${\em Service de Physique Th\'eorique et Math\'ematique,  Universit\'e Libre de
Bruxelles,
\\ Campus Plaine C.P. 231\\ Boulevard du Triomphe, B-1050 Bruxelles, Belgium}\\ {\tt
lhouart@ulb.ac.be}

\vspace{.2cm}

${}^c${\em The International Solvay Institutes, \\Campus Plaine C.P. 231\\ Boulevard du
Triomphe, B-1050 Bruxelles, Belgium}\\

\end{center}

\vspace {2cm}
\centerline{ABSTRACT}

The set of exact solutions of the non-linear realisations of the $\G$ Kac-Moody
algebras is further analysed. Intersection rules for extremal branes translate 
into orthogonality conditions on the positive real roots characterising each brane.  It is
proven that all the intersecting extremal brane solutions of the maximally oxidised
theories have their algebraic counterparts as exact solutions in the $\G$-invariant
theories. The proof is extended to include the  intersecting extremal brane
solutions of the  exotic phases of the maximally oxidised theories.

\vspace{- 3mm}
\begin{quote}\small

\end{quote}

\newpage
\baselineskip18pt

\setcounter{equation}{0}
\addtocounter{footnote}{-1}

\section{Introduction}

A maximally oxidised theory associated with a simple group $\cal G$ is a theory of 
gravity coupled to forms and dilatons defined at the highest possible space-time
dimension $D$ which upon dimensional reduction down  to three  is expressible in terms
of a coset space ${\cal G}/{\cal H}$ where
$\cal H$ is the maximally compact subgroup of $\cal G$. The maximally oxidised actions
$S_{\cal G}$ corresponding to all the simple group $\cal G$ have been classified
\cite{cjlp}  and they comprise in particular pure gravity in $D$ dimensions, the bosonic
part of the low energy effective action of M-theory and the low energy effective action
of the bosonic string. It has been conjectured that these actions, or some extension of
them,  possess the much larger very-extended Kac-Moody symmetry
$\G$. $\G$ algebras are defined form the Dynkin diagrams obtained from those
of $\cal G$ by adding three nodes
\cite{ogw}. One first adds the affine node, then a second node connected to it by a single
line to define the overextended ${\cal G}^{++}$ algebras,  then similarly a third one
connected to the second to define the very-extended $\G$ algebras. Such $\G$
symmetries were first conjectured in the above mentioned particular case
\cite{west01,lw} and the extension to all 
$\G$ was proposed in 
\cite{ehtw}. In a different development, the study of the properties of cosmological
solutions in the vicinity of a space-like singularity, known as cosmological billiards
\cite{damourhn00}  revealed an  overextended symmetry ${\cal G}^{++}$ for all $\cal G$
\cite{damourh00, damourbhs02}.

Motivated by these developments and by the approach to $E_8^{++}$ proposed in
reference \cite{damourhn02}, we  formulated in
\cite{eh}  an explicit  non-linear realisation for all simple $\G$. These actions $\cal S_{\cal
G}$ are defined with  no a priori reference to space-time, which is expected to be
generated dynamically. These
$\G$-invariant actions are proposed as substitutes for the original field theoretic
models of gravity, forms and dilatons and hopefully contain new degrees of freedom such
as those encountered in string theories. They are formulated recursively from a level 
decomposition \cite{damourhn02, west02, west05} with respect to a subalgebra
$A_{D-1}$ where $D$ turns out to be the space-time dimension. The fields appearing in the
actions live in a coset space $\G /K^{+++}$ where the subalgebra
$K^{+++}$ is invariant under a `temporal involution' \cite{eh} which is different from the
often used Chevalley involution. The temporal involution preserves the Lorentz algebra
$SO(1,D-1)$ and as a consequence the actions $\cal S_G$ are Lorentz invariant at each
level. Exact solutions of these $\G$-theories describing the algebraic properties of the 
extremal branes of the corresponding  maximally oxidised theories have been obtained 
\cite{eh}, and the existence of `dualities' for all $\G$ have been traced to  their
group-theoretical origin. The `dualities' are described by  Weyl reflections in
$\G$.

Here  we  extend these results. In Section 2 we find exact solutions of $\G$ corresponding
to  {\it all} the   extremal intersecting brane solutions of the maximally oxidised
theories. To establish this result, we first show that the extremal  branes and  the 
intersection rules \cite{aeh}  characterising their intersections are neatly encoded  in 
the $\G$ algebras. Namely each extremal brane corresponds to a real positive root of the
$\G$ algebra and the intersection rules for the branes translate into orthogonality
conditions on their roots. These results permit a truncation of the actions $\cal S_G$ to
their quadratic expansion from which the exact solutions are derived. In Section 3 these
results are further extended for all $\G$ to `exotic' phases, which may have different
space-time signature. It has indeed been shown in
\cite{keu} that the `temporal' involution is not invariant under all the Weyl reflections. 
In
the particular case of $E_8^{+++}$ this implies that the exotic phases of M-theory
\cite{compt,exop} reached by timelike $T$-duality in the  string language are included in
the formalism of \cite{eh}. In general
 exotic phases for all $\G$ are also included.

\section{Intersecting brane configurations as exact solutions in ${\cal G}^{+++}$}

In \cite{eh}, theories invariant under $\G$  where constructed and exact  solutions
describing the algebraic properties of  the   BPS extremal brane solutions of all the
 maximally oxidised theories  associated with the simple groups $\cal G$ 
were presented. In this section, we  extend these results. We proof   that all the
intersecting extremal  brane solutions of these theories have also their algebraic 
counterpart as exact solutions in $\G$.  
As an element of this general proof we find the elegant encoding in  the $\G$
algebras of the intersections.

In order to establish this result we first  analyse  the relations between brane
dynamics and symmetry in the maximally oxidised  theories.  We first
recall the generic intersection rules \cite{aeh} which determine how extremal branes can
intersect orthogonally  with  zero binding energy. These intersection
rules are valid for a generic theory in $D$ dimensions which includes gravity, a dilaton, 
form field strengths $F_{n_I}$ of arbitrary degree $n_I$  and arbitrary couplings to the
dilaton $a_I$
\footnote{We consider $n_I \leq D/2$. If $n_I >D/2$, we can indeed replace the field
strength we start with by its Hodge dual.}. They give for each pair
$(A,B)$ of $q$-branes of dimensions
$(q_A,q_B)$, the number of dimensions $\bar q$ on which they intersect in  terms of the
total number of space-time dimensions $D$ and of the field  strength couplings to the
dilaton. They read \cite{aeh}
\begin{equation}
\bar{q}+1=\frac{(q_A+1)(q_B+1)}{D-2}-\frac{1}{2}
\varepsilon_A a_A \varepsilon_B a_B \, ,\label{intrule}
\end{equation} where $\bar{q}$ is  the number of spatial dimensions on which the $q_A$
and the $q_B$ brane intersect, $\epsilon_A$ is $+1$ (resp. $-1$) if the $q_A$-brane is 
electric (resp. magnetic)\footnote{ The case $\bar{q}=-1$ is also relevant and have an interpretation in 
terms of instanton in the Euclidean. Then ,the time coordinate  doesn't need to be
longitudinal to all the branes.}. An intersecting extremal brane configuration exists thus
between  two branes if $\bar q$ is an integer  not bigger than  the  brane of lowest
dimension.  We restrict ourself to intersecting
brane configurations characterised by a space which is asymptotically flat, namely  we
consider configurations with an overall transverse space $d \geq 3$. In the derivation of
Eq.(\ref{intrule}), it is assumed that
in the configuration considered there are no contributions to the equations of
motion from the Chern-Simons terms that can be present in the action.

We are interested in intersecting brane configurations of the maximally oxidised
theories corresponding to any simple group $\cal G$ \cite{cjlp} and characterised by at
most one dilaton\footnote{We are  considering  all the maximally oxidised theory
except the ones corresponding to  the $C_{q+1}$-series. The maximally oxidised
theory $C_{q+1}$ is a four dimensional theory which contains $q$-dilatons.}. In
ref.\cite{ehw} it
has been found that the scalar products between first electric and magnetic roots
encountered in the dimensional reduction process and corresponding to any of  the form 
field strengths are given by the intersection rules. 
More precisely one gets the following relations. First one has  \cite{ehw}
\begin{eqnarray} 
\alpha_{A}^e \cdot\alpha_{B}^e  &=& {\bar q}^{(e_A,m_B)} +1\label{ee1}\\ &=&
q_A^e-{\bar q}^{(e_A,e_B)}\, ,\label{ee2}
\end{eqnarray} where $A$ (resp.$B$) refers to the form field strength $F_{n_A}$ (resp.
$F_{n_B}$) present in the maximally oxidised  $\cal G$ theory, and  $n_A \leq  n_B$. The 
$\alpha_X^e$ with $X=(A,B)$ is the first electric root coming from the 
$F_{n_X}$ upon dimensional reduction, that is the one appearing  when one reaches
$n_X-1$   compact dimensions. The  superscripts of $\bar q$ label the electric or
magnetic nature of the $q_A$-brane and  the $q_B$-brane. One has
$q_X^e= n_X-2$ and
$q_X^m=D-q_X^e-4$. Using these relations, we  express in Eq.(\ref{ee2}) the
scalar product  of the two first electric roots in terms of the intersection between the
two   corresponding electric branes. 
Second, one gets for the scalar products  between a first electric root and a first magnetic
root
\begin{eqnarray} 
\alpha_{A}^e \cdot \alpha_{B}^m  &=& {\bar q}^{(e_A,e_B)} +1 \label{em1}\\ &=&
q_A^e-{\bar q}^{(e_A,m_B)}\, ,\label{em2}
\end{eqnarray}
with again $n_A \leq n_B$.
The first magnetic root $\alpha_X^m$ corresponds in the dimensional reduction to the
additional scalar coming from $F_{n_X}$ which arises when one reaches $n_X+1$
non-compact dimensions and is obtained by dualizing the $n_X$-form. 
We have expressed the scalar product in
terms of the intersection of $q_A^e$ and $q_B^m$ (note that since in our setting $n_X
\leq D/2$, one has always $q_X^e \leq q_Y^m$).  Third, the scalar product between two first
magnetic roots is given by
\begin{eqnarray} 
\alpha_{A}^m \cdot\alpha_{B}^m  &=& {\bar q}^{(e_A,m_B)} +1 \label{mm1}\\ &=&
q_B^m-{\bar q}^{(m_A,m_B)}\label{mm2}\, .
\end{eqnarray} Here $q_B^m \leq q_A^m$. Thus  the smallest brane
 appears always  in the r.h.s of Eqs.(\ref{ee2}), (\ref{em2}) and (\ref{mm2}).

We now use these results in the context of the actions proposed in \cite{eh}  which are
explicitly invariant under the very-extended algebras $\G$.

We  recall how the $\G$-invariant actions $\cal S_G$ were constructed   recursively  from
a level expansion with respect to a subalgebra
$A_{D-1}$ where $D$ is  the space-time dimension \cite{eh}. At each level the
$SO(1,D-1)$ invariance is  realised through the use of the `temporal' involution instead of
the  usual Chevalley one in the construction of the non-linear  realisation.
$\G$ contains a  subalgebra $GL(D)$ and we have $SL(D) \subset GL(D) \subset \G$. The
generators of the $GL(D)$ subalgebra are taken to be 
$K^a{}_b\ (a,b=1,2,\ldots ,D)$   with commutation relations
\begin{equation}
\label{Kcom} [K^a_{~b},K^c_{~d}]   =\delta^c_b K^a_{~d}-\delta^a_dK^c_{~b}\,  .
\end{equation}  The $K^a{}_b$ along with the the abelian generator $R$ (present  when the
corresponding $\cal G$ theory has one dilaton) are the level zero generators. The
operators of level greater than zero are tensors of the 
$A_{D-1}$ subalgebra. The lowest levels contain antisymmetric tensor step  operators
$R^{a_1a_2 \dots a_r}$ associated with the electric and magnetic  roots occurring in the
dimensional reduction of the corresponding maximally oxidised theory. They satisfy
\begin{eqnarray}
\label{root} &&[K^a_{~b},R^{a_1\dots a_r}]   =\delta^{a_1}_b R^{aa_2 \dots a_r} +\dots +
\delta^{a_r}_b R^{a_1 \dots a_{r-1}a}\, ,\\
\label{root2} && [R,R^{a_1\dots a_r}] =   -\frac{\varepsilon_A a_A}{2}\,  R^{a_1\dots
a_r}\, ,
\end{eqnarray} where $a_A$ is the dilaton coupling of the corresponding form field
strength in the $\cal G$ theory and $\varepsilon_A$ is $+1\, (-1)$ if the corresponding 
root is electric (magnetic). One defines fields in a one dimensional space $\xi$, a priori
unrelated to space-time, as the  parameters of the group elements $\cal V$  built out of
Cartan and positive step operators in $\G$. It takes the form
\begin{equation}
\label{positive} {\cal V}= \exp (\sum_{a\ge b} h_b^{~a}(\xi)K^b_{~a} -  
\phi(\xi) R) \exp (\sum
\frac{1}{r!s!} A^{\quad a_1\dots a_r}_{ b_1\dots b_s}(\xi) R_{\quad a_1\dots   a_r}^{
b_1\dots b_s} +\cdots)\, ,
\end{equation}
 where the first exponential contains only  level zero   operators  and the second one the
positive step operators of level strictly greater than zero. In terms of these fields, the 
action $\cal S_G$ is 
\begin{equation}
\label{full}  {\cal S_G}={\cal S_G}^{(0)}+\sum_A{\cal S_G}^{(A)}\, ,
\end{equation} where ${\cal S_G}^{(0)}$ contains all level zero contributions. Explicitly one
has
\begin{equation}
 {\cal S_G}^{(0)}=\frac{1}{2}\int d\xi
\frac{1}{n(\xi)}\left[\frac{1}{2}(g^{\mu\nu}g^{\sigma\tau}- 
\frac{1}{2}g^{\mu\sigma}g^{\nu\tau})\frac{dg_{\mu\sigma}}{d\xi}
\frac{dg_{\nu\tau}}{d\xi}+
\frac{d\phi}{d\xi}\frac{d\phi}{d\xi}\right]
\end{equation}
$${\cal S_G}^{(A)}=\frac{1}{2}\int d\xi
\frac{1}{n(\xi)}\left[\frac{1}{r!s!}\exp (- 2\lambda
\phi)
\frac{DA_{\mu_1\dots \mu_r}^{\quad \nu_1\dots
\nu_s}}{d\xi} g^{\mu_1{\mu}^\prime_1}...\,
g^{\mu_r{\mu}^\prime_r}g_{\nu_1{\nu}^\prime_1}...\,    g_{\nu_s{\nu}^\prime_s}
\frac{DA_{{\mu}^\prime_1\dots {\mu}^\prime_r}^{\quad {\nu}^\prime_1\dots
{\nu}^\prime_s}}{d\xi}\right]$$ Here, $g_{\mu\nu} =e_\mu^{~a}e_\nu^{~b}\eta_{ab}$,  
$e_\mu^{~a}=(e^{-h(\xi)})_\mu^{~a}$ with $(a,b)$  Lorentz indices and $(\mu,\nu)$
$GL(D)$ indices, $\lambda$ is the generalization of $-\varepsilon_A a_A/2$ to all roots
and $D/D\xi$ the non-linear covariant derivative generalising  $d/d\xi$ to take into
account  non vanishing commutators  between  positive level step operators. The
arbitrary lapse function $  n(\xi)$ renders
$\cal S_G$ reparametrisation invariant.

The exact solution of the $\G$ theories \cite{eh}  corresponding to a single extremal
$q$-brane longitudinal to the $\lambda_1 \dots \lambda_q$ spatial directions  is always
electrically described\footnote{If the brane is a magnetic one, it is described by the
Hodge dual field whose potential appear as well in  the low levels. The time  direction is
taken to be the direction 
${\hat 1}$ thus  $x^1=t$.} and is characterised by only  one non-zero field component 
 $A_{t\lambda_1 \dots
\lambda_q}$. This field is the parameter of an antisymmetric tensor  step operator of low
level 
$R^{1 \lambda_1 \dots
\lambda_q}$ and we denote the  corresponding  {\it real positive root}\footnote{In
ref.\cite{bgh}, they argued   in the context of $E_{10}=E_8^{++}$ that extremal branes
correspond to some imaginary roots. It is worthwhile to emphasise that  in the present
$\G$ framework {\it all}  the extremal branes correspond to real positive  roots.} by
$\alpha_{(1,\lambda_1, \dots ,\lambda_q)}$.

Suppose now that we have two branes, one extremal $q_A$-brane along the 
$\lambda_1 \dots \lambda_{q_A}$ spatial directions  associated  with the root
$\alpha_{(1,\lambda_1, \dots ,\lambda_{q_A})}$  and an extremal $q_B$-brane with $q_B
\geq q_A$ along the $\nu_1 \dots \nu_{q_B}$ spatial directions associated with the root
$\alpha_{(1,\nu_1, \dots ,\nu_{q_B})}$ . We assume that the branes  have ${\bar q}$ indices
in common namely $\lambda_i=\nu_i$ for $\bar q$  different $i$'s. We want first to
demonstrate the following statement which translates in a group language the
intersection rules

\noindent{\bf Theorem 1:} The existence of an integer solution ${\bar q} \leq q_A$  of the
intersection rule equation between two extremal branes  Eq.(\ref{intrule}) is equivalent
to  the following condition\footnote{This condition has been noticed, in a somewhat
different setting, for some particular intersecting configurations of M-theory in
\cite{westxx}.}  on the two real positive roots corresponding to  the two branes
\begin{equation}
\alpha_{(1,\lambda_1, \dots ,\lambda_{q_A})} \cdot 
\alpha_{(1,\nu_1, \dots ,\nu_{q_B})} = 0\, .
\label{cond1}
\end{equation}

In order to show that we will use the relations between first electric and magnetic roots
derived in the context of the dimensional reduction. We first note that the scalar
product  between two roots is invariant under Weyl reflection and that the Weyl
reflections generated by the simple roots $\alpha_i^g, \quad i=1 \dots D-1$  of the gravity line
defined by the subalgebra $A_{D-1}$
simply exchanges $x^i \leftrightarrow x^{i+1}$ (including time $t=x^1$). 
Using Weyl reflections, we can thus  map 
$\alpha_{(1,\lambda_1, \dots ,\lambda_{q_A})}$ onto $\alpha_{(D-q_A,D-q_A+1, \dots ,D)}$
 and  map $\alpha_{(1,\nu_1, \dots ,\nu_{q_B})}$ onto 
$\alpha_{(D-q_A-q_B+{\bar q},  \dots ,  D-q_A-1,D-{\bar q}, \dots ,D)}$. These roots share
the directions $D-{\bar q}, \dots ,D$. We have
\begin{eqnarray}
\alpha_{(1,\lambda_1, \dots ,\lambda_{q_A})} \cdot 
\alpha_{(1,\nu_1, \dots ,\nu_{q_B})} &=&
\alpha_{(D-q_A,D-q_A+1, \dots ,D)} \cdot 
\alpha_{(D-q_A-q_B+{\bar q},  \dots ,  D-q_A-1,D-{\bar q}, \dots ,D)}
\nonumber\\ &\equiv& \alpha_A \cdot \alpha_B^{\prime}\label{cond2}\, .
\end{eqnarray} Since we made the  assumption that in the intersecting brane
configurations  the overall transverse space is $d \geq 3$, we have
$D-q_A-q_B+{\bar q} \geq 4$ and $\{\alpha_A  ,  \alpha_B^{\prime}\} \in {\cal G}
\subset \G$. The root $\alpha_A$  corresponds to the generator $R^{D-q_A \dots D}$ and,
in the dimensional reduction language is a first electric (or magnetic) root. The root
$\alpha_B^{\prime}$ corresponds to the generator 
$R^{D-q_A-q_B+{\bar q} \dots  D-q_A-1 ~ D-{\bar q} \dots D}$ and it is not a first
electric or magnetic root. In order to be able to use the results
Eqs.(\ref{ee2}),(\ref{em2}) and (\ref{mm2}) to compute $\alpha_A \cdot
\alpha_B^\prime$ we need to express the root $\alpha_B^\prime$ in terms of the simple
roots of the gravity line and in terms of the  first root $\alpha_B$
corresponding to the generator $R^{D-q_B D-q_B-1 \dots D}$. We can relate 
$R^{D-q_A-q_B+{\bar q} \dots  D-q_A-1 ~ D-{\bar q} \dots D}$ to $R^{D-q_B D-q_B-1
\dots D}$ with  Eq.(\ref{root}). Using the fact that the simple roots $\alpha_i^g$ of the
gravity line correspond to the generators $K^{i}_{~i+1}$, one finds the following relations
between $\alpha_B$ and $\alpha_B^\prime$
\begin{equation}
\alpha_B^\prime = \alpha_B+ (q_A- {\bar q}) \alpha_{D-q_a-1}^g + \Lambda
\label{bbprime}\, ,
\end{equation} where $\Lambda$ is a sum of simple roots of the gravity line with positive
integer coefficients which {\it does not contain} a contribution from
$\alpha_{D-q_a-1}^g$. We have thus
\begin{equation} 
\Lambda \cdot \alpha_A=0\, .
\label{Lambda}
\end{equation}  Now using  Eqs.(\ref{ee2}),(\ref{em2}) and (\ref{mm2}) together with
Eqs.(\ref{bbprime}) and (\ref{Lambda}),   we find
\begin{eqnarray}
\alpha_A \cdot \alpha_B^\prime &=& \alpha_A.\alpha_B + (q_A -{\bar q})
\, \alpha_A \cdot \alpha_{D-q_A-1}^g \nonumber\\ &=& q_A-{\bar q} +(q_A-{\bar q})
(-1)\nonumber\\ &=& 0 \label{claim1}\, ,
\end{eqnarray} which demonstrates theorem 1.

We are now in position to construct exact solutions of the $\G$
actions corresponding to extremal intersecting brane solutions. 
It will be sufficient for the construction to analyse pairwise intersections.
Consider two branes, one
corresponding to the root
$\beta \equiv \alpha_{(1,\lambda_1, \dots ,\lambda_{q_A})}$ and the other 
 to $ \gamma \equiv \alpha_{(1,\nu_1, \dots ,\nu_{q_B})}$.
The branes  have ${\bar q}$ indices in common namely
$\lambda_i=\nu_i$ for $\bar q$  different $i$'s.  We label the two roots such  that the level
of $\gamma$ is not lower than the level of $\beta$. We also lift the restriction that the
overall transverse space $d \geq 3$, namely we admit intersecting brane configurations
which are  no longer necessarily in a $\cal G \subset
\G$. We claim

\noindent {\bf Theorem 2:} There exists a solution of the $\G$ action describing the 
intersection of two extremal branes  associated with the real positive roots $\beta$ and $\gamma$ iff
\begin{eqnarray} & \beta \cdot \gamma = 0
\label{conda}\\ &{\rm and} \nonumber\\ & \beta+ 
\gamma \not= {\rm root}\,  .\label{condb}
\end{eqnarray}

To establish theorem 2 we will use the following lemma

\noindent {\bf Lemma:} if  Eq.(\ref{conda}) is  satisfied then Eq.(\ref{condb}) is equivalent
to
\begin{equation}
\gamma \not=
\beta +{\tilde \alpha}\, ,\label{condc} 
\end{equation} where $\tilde \alpha$ is any positive root of $\G$.

To proof the lemma we first  show, using Eq.(\ref{conda}) and Eq.(\ref{condb}), that
$\gamma - \beta$ is not a root namely that $[E_\gamma, F_\beta]=0 $ where
$F_\beta$  is the step operator corresponding to the negative root $-\beta$. This is a
consequence of the Jacobi identity
$[ F_\beta [E_\gamma, E_\beta]]+[E_\gamma[E_\beta, F_\beta]]+[E_\beta[F_\beta,
E_\gamma]]=0$. The first term is zero because of Eq.(\ref{condb}). To evaluate the
second term we use the standard invariant bilinear form $K$ defined on a Kac-Moody
algebra \cite{kac83}. We consider the following identity reflecting the invariance of
$K$: 
$K([E_\alpha, F_{\alpha}],H_n)+ K(F_\alpha, [E_\alpha, H_n])=0$ where $H_n$ is the Cartan
generator in the Chevalley basis corresponding to the simple root $\alpha_n$. Writing
$\alpha$ as a sum of simple roots $\alpha_m$: 
$\alpha = \sum_{m=1}^r N_m \alpha_m$ where $N_m$ are non-negative integers and
using the fact that for $\G$ the determinant of the Cartan matrix is different from zero, 
we deduce from the above identity  that
$[E_\alpha, F_\alpha]= (1/2) K(E_\alpha, F_\alpha)\sum_m N_m \alpha_m^2 \, H_m$. From
there it follows that  $[E_\gamma, [E_\beta, F_\beta]]= \,-K(E_\beta,F_\beta) \,
(\beta \cdot \gamma) \, E_\gamma$. Thus this term is also zero by 
Eq.(\ref{conda}).  Furthermore, since $\gamma$ is a root different from $\beta$, $
[E_\beta[F_\beta E_\gamma]]=0$ only if 
$[F_\beta E_\gamma]=0$ thus Eq.(\ref{condb}) $\Rightarrow$ Eq.(\ref{condc}). Using 
Eq.(\ref{conda}) and the same Jacobi identity one shows also that Eq.(\ref{condc})
$\Rightarrow$ Eq.(\ref{condb}). In that case  the second and the third  term are zero and
the first one implies that $\beta + \gamma$ is not a root. This concludes the proof of the
lemma.

An exact solution of  $\cal S_G$  corresponding to an intersecting 
brane configuration is constructed as a  generalisation of
 single extremal brane solutions discussed in \cite{eh}. For each of the $\cal N$ branes
present in the configuration the solution has a  non-zero field component given by
\begin{equation} A_{t\lambda_1\dots \lambda_{q_A}}=
\epsilon_{t\lambda_1\dots\lambda_{q_A}} [\frac{2(D-2)}{\Delta_A}]^{1/2}H_A^{-1}(\xi)\, ,
\qquad A=1 \dots {\cal N}\, ,
\label{aequ}
\end{equation} and, defining $p^{(a)}= p^{(\mu)}= \ln  e^a_\mu$ for the diagonal vielbein
in a triangular gauge, dilaton and metric components given by
\begin{eqnarray}
 &&p^{(\mu)}= \sum_{A=1}^{\cal N} p_A^{(\mu)}=\sum_{A=1}^{\cal N} 
\frac{\eta_A^{\mu}}{\Delta_A}  
\ln H_A(\xi)~  \label{pequ}\\
\label{phiequ}
&&\phi =\sum_{A=1}^{\cal N} \, \phi_A = \sum_{A=1}^{\cal
N}\frac{D-2}{\Delta_A}\varepsilon_A a_A 
\ln H_A(\xi) \, .
\end{eqnarray} Here $\eta_A^{\mu}=q_A+1$ or $-(D-q_A-3)$ depending on whether 
the direction $\hat{\mu}$ is perpendicular or parallel to the $q_A$-brane and 
$\Delta_A= (q_A+1)(D-q_A-3)+\frac{1}{2}a_A^2(D-2)$. Each of the  branes in the
configuration is characterised by one harmonic function in 
$\xi$-space, namely one has 
\begin{equation}
\frac{d^2H_A(\xi)}{d\xi^2}=0 \qquad A=1 \dots {\cal N}\, .
\label{harmo} 
\end{equation} 

In order to show that Eqs.(\ref{aequ})-(\ref{harmo}) constitute indeed a solution of the
equations of motion  we proceed in two steps.

First we make the working hypothesis that we can substitute $\cal S_G$ by its
 quadratic truncation, as we did  for the single brane solution in reference \cite{eh}. The
quadratic action is defined by  expanding $\cal S_G$ given in Eq.(\ref {full}) in power of
fields   parametrizing the positive step operators up to quadratic terms.
Under this  assumption, we check  that Eqs. (\ref{aequ})-(\ref{harmo}) are
solutions of the equations of motion (equations (3.10)-(3.14) of ref.\cite{eh}).
One potential problem    arises from the argument of  the exponential in front of a  
$dA_{t
\lambda_1 \dots \lambda_{q_A}} /d\xi$ term  associated with a  $q_A$-brane
present in the configuration : this argument is given by 
$\varepsilon a \phi -2p^{(t)}-2\sum_{\lambda=\lambda_1}^{\lambda_r}
p^{(\lambda)} $.  We verify however that the contribution to this expression  of any other
$q_B$-brane present in the configuration vanishes identically. This is 
easily shown using the intersection rule Eq.(\ref{intrule}) between $q_A$ and
$q_B$, which is equivalent, as shown in theorem 1, to Eq.(\ref{conda}). 
Note that it follows from the above solution and from the intersection rules that the 
lapse constraint is satisfied and takes the form
\begin{equation}
\label{xiextremal}
\sum_{\alpha=1}^D ( d p^{(\alpha)})^2 -\frac{1}{2}(\sum_{\alpha=1}^D
dp^{(\alpha)})^2 +
\frac{1}{2} (d\phi)^2-  \sum_{A=1}^{\cal N}\frac{D-2}{\Delta_A}(d\ln H_A)^2=0\, .
\end{equation}

Provided the truncation of the $\cal S_G$ actions to their quadratic form is  consistent
 we have thus an exact solution.  The conditions Eq.(\ref{condb}) and
Eq.(\ref{condc})  are precisely those ensuring that the replacement of the actions by
their quadratic simplified versions  is consistent. 
First, Eq.(\ref{condb}) ensures that the substitution of the solution
Eqs.(\ref{aequ})-(\ref{harmo}) in the action $\cal S_G$ leads only to quadratic terms.
Second, we check that for the field components $\tilde A$ which are zero in the
configuration, $\tilde A=0$ is solution of  the equations of motion  of the full $\cal S_G$.
This is the case.
 Indeed,  on the one hand, Eq.(\ref{condb}) ensures
that the  covariant derivative of $\tilde A$ does not contain non-linear terms built
purely  out of non-zero field components in the configuration.  On the other hand, 
Eq.(\ref{condc}) ensures that in the  covariant  derivative of a non-zero $A$ field
component in the configuration there are no non-linear terms  which would contain
other non-zero
$A$ field components in the configuration along with a $\tilde A$ field
component.
The equations of
motion of the $\tilde A$'s are thus  trivially satisfied putting these $\tilde A$'s to zero.
This concludes the proof  of theorem 2.  

In addition to the above considered charged extremal brane,  there exist  two other
gravitational  BPS branes in the $\G$ theories, namely the KK-momentum and the
KK-monopole. These are related by Weyl reflections to the charged BPS branes \cite{eh} 
and since  Weyl reflections preserve the scalar product, theorem 2   extends to
configurations containing also gravitational branes. The root of a
KK-momentum in the
$x^k$ direction is the one associated with the positive level zero step  operator
$K^1_{~k}$ and  the root  of the KK-monopole with, say, the longitudinal
directions
$(x^2, \dots , x^{D-4})$ and Taub-NUT direction $x^D$ is the one associated with the
positive step operator $R^{1 \dots D-4D,D}$  (see Appendix B of \cite{eh}).  This step
operator, which is  antisymmetric in the first $D-3$ indices and  with a vanishing totally
antisymmetrised contribution, exists   at some level for all $\G$
\cite{west05}.

We now turn to our central theorem

\noindent {\bf Theorem 3:}  There is a one to one correspondence between  the  exact
solutions of 
$\cal S_G$ 
given by  theorem 2 and  the intersecting  extremal brane solutions of the 
maximally oxidised theory $S_{\cal G}$.

We first proof the theorem for $\cal G$ simply laced.
First recall that the intersection rules Eq.(\ref{intrule})  characterising the intersecting
brane solutions of the maximally oxidised theories have  been derived under the
assumption that in the given configuration the Chern-Simons  terms do not contribute to
the equations of motion. This is the case for all the phases of M-theory ($11D$ SUGRA,
$IIA$ and $IIB$) \cite{aeh} . More generally one can check by inspection  of the
explicit form of the  actions \cite{cjlp}  that for all maximally oxidised theory
corresponding to simply laced group $\cal G$, it is also the case.
Second, we note that for simply laced $\G$ theories theorem 2
simplifies.  In that case, Eq.(\ref{condb}) and thus also Eq.(\ref{condc}) are trivially
satisfied once the Eq.(\ref{conda}) is implemented. Indeed, for simply laced theories,  all
the real roots
$\alpha^R_i$ have the same length  (say $(\alpha^R_i)^2=2$) thus if
$\alpha^R_i \cdot \alpha^R_j =0$, $ \alpha_+ \equiv \alpha^R_i+\alpha^R_j$ can not be a
root because
$\alpha_+^2=4$ and Eq.(\ref{condb}) is satisfied.
Consequently,  Eq.(\ref{condb}) and Eq.(\ref{condc}), eliminating potential problems
related to the non-linear terms in the equations of motion of $\cal S_G$,   are trivially
satisfied. Thus the intersection rule, which in $\G$ is implemented by Eq.(\ref{conda}),
fixes uniquely the intersecting brane solutions, both in $S_{\cal G}$ and in $\cal S_G$.
This establish the proof of theorem 3 for simply laced $\cal G$.

We now turn to the  non-simply laced theories with one dilaton namely the
$B_{D-2}$  series and $F_4$ ($G_2$ does not admit intersecting brane solutions).

We discuss first  in detail the $B_{D-2}$ series. The maximally oxidised 
${\cal G}=B_{D-2}$  theory in $D$-dimensions with one dilaton  contains a three-form field
strength $F_3$  with dilaton coupling $a_3=-l$ and a two-form  field strength
$F_2$ with dilaton coupling $a_2=-l/2$ where 
$l^2=8/(D-2)$. There is an electric extremal $0$-brane  and a magnetic $(D-4)$-brane
associated with $F_2$ and an electric
$1$-brane and a magnetic $(D-5)$-brane associated with $F_3$. In  ten dimension it
corresponds to the bosonic part of the  low energy effective  action of the heterotic
string truncated to only one gauge field. The dilaton couplings are such that the
intersection rules  Eq.(\ref{intrule}) predict among the possible intersecting brane
configurations the following ones associated with $F_2$: 
$0 \cap 0 =-1$,
$0 \cap (D-4)=0$ and  $(D-4) \cap (D-4)=D-5$.  Inspecting the action of the maximally
oxidised theory $B_{D-2}$ given in \cite{cjlp}, it is easy to see there is a Chern-Simons
like term appearing in the Bianchi identity $dF_3= (1/2)\,  F_2
\wedge F_2$, which does not vanish in these three configurations. Consequently these
intersections are not solutions of the maximally oxidised theory and  do not exist. The
other possible intersecting brane configurations predicted by Eq.(\ref{intrule}) are not
invalidated by  such  Chern-Simons like term and do exist.

In the $\G=B_{D-2}^{+++}$ theory these three configurations are also discarded because 
they violate  Eq.(\ref{condb}) of theorem 2 (or equivalently Eq.(\ref{condc})). To see that,
we recall the level decomposition of $B_{D-2}^{+++}$ in terms of
$A_{D-1}$\cite{west05}. There are two simple roots which do not belong to the gravity
line. The first one
$\alpha_D$ is short and corresponds  to $R^{D}$ associated with the component
$A_{D}$ of the one-form potential. The second root not belonging to the gravity line
$\alpha_{D+1}$ is a long one and corresponds to $R^{5 \dots D}$ associated with the
component $A_{5 \dots D}$ of the $(D-4)$-form potential. The level decomposition in terms
of $A_{D-1}$  representations is thus labelled by two non-negative integers $(l_1, l_2)$
giving the number of time the two roots $(\alpha_{D+1}, \alpha_D)$ appear in a given
representation. The lowest level corresponding to the different potentials associated
with branes  are the following \cite{west05}
\begin{center}
\begin{tabular}{|lll|}
\hline level & $\alpha^2$ & potential\\
\hline
$(0,1)$ & 1 & $A_1$\\
$(0,2)$ & 2 & $A_2$\\
$(1,0)$ & 2 & $A_{D-4}$\\
$(1,1)$ & 1 & $A_{D-3}$\\
\hline
\end{tabular}
\end{center} Now it is easy to see that the three configurations are eliminated. The
configuration $0 \cap 0=-1$ does not satisfy Eq.(\ref{condb})   indeed the sum of the two
roots corresponding to the two $0$-branes is a root of level $(0,2)$.  For the
configuration $(D-4) \cap 0=0$ it is  easier to see that Eq.(\ref{condc}) is violated. 
Indeed
the root of  level $(1,1)$ corresponding to the $(D-4)$-brane is the sum of the root of level
$(0,1)$  corresponding to the $0$-brane and of a root of level $(1,0)$ corresponding to the
potential $A_{y_1 \dots y_{D-4}}$ where the $y_i$ are the spatial longitudinal
coordinates of the $(D-4)$-brane. Finally, as far as the configuration
$(D-4) \cap (D-4) =D-5$ is concerned, we need to know the level decomposition at higher
levels. 
At level $(2,2)$ there is a representation \cite{west05} corresponding to  real  long
roots characterised by the following Dynkin labels
\footnote{Here we follow the usual convention. The last label on the right refers to the
fundamental weight associated with the `time' root.} : $(1,0 \dots 0,1,0)$. The sum of the 
roots corresponding to the two $(D-4)$-brane is a long root belonging to that
representation. Consequently this configuration  is not solution.

We thus conclude that   each time there is an intersecting brane solution in the maximally
oxidised
$B_{D-2}$ theory, its algebraic counterpart exists as an exact solution of 
$B_{D-2}^{+++}$ theory. The solutions predicted by the intersection rules
Eq.(\ref{intrule}) which are eliminated because of  Chern-Simons terms in the maximally
oxidised theory are also eliminated in
$B_{D-2}^{+++}$ because of the existence of  non-linear terms  in the configuration.
Furthermore the intersections between pair of branes which are discarded
correspond always to configuration containing two branes, magnetic or electric,
corresponding to two short roots.

A similar  discussion can be performed in the $F_4$ case.  The maximally oxidised theory
${\cal G}=F_4$ is a six dimensional theory with one dilaton, a one-form field strength, two
two-form field strengths and two three-form field  strengths
\cite{cjlp}. There are a lot of  possible intersecting brane pairs in this theory. It suffices
to consider the configurations  involving two  extremal branes corresponding to two 
short roots. They are  associated  with several field strengths,  the two two-form field 
strengths
$F_2$ and
$F_2^\prime$ with dilaton couplings $a_2=-1/ \sqrt{2}$ and $a_2^\prime=1/ \sqrt{2}$,
the one-form field strength $F_1$ with dilaton coupling $a_1= \sqrt{2}$ and the self-dual
three-form field strength $F_3$.  To
each two-form corresponds an electric $0$-brane and a magnetic $2$-brane, a $(-1)$-brane 
and a $3$-brane are associated with $F_1$  and a self-dual $1$-brane is associated with  $F_3$.
The intersection rules predict the following configurations between these electric and magnetic branes: 
$0 \cap 0 =-1$, $0 \cap 2=0$, $2 \cap 2 =1$,
$0^\prime \cap 0^\prime =-1$, $0^\prime \cap 2^\prime=0$,  $2^\prime \cap 2^\prime=1$,
$1 \cap 1 =0$, $-1 \cap 1 =-1$, $3 \cap 1=1 $. These intersections predicted by
Eq.(\ref{intrule}) are not solutions of the maximally oxidised theory. Indeed  as  for the
$B_{D-2}$, there are Chern-Simons type terms (see \cite{cjlp}) which are non zero in these
configurations. The other intersections predicted by Eq.(\ref{intrule}) are solutions.
 Again, these configurations are also eliminated in the $\G =F_4^{+++}$ theory.
Using the level decomposition of $F_4^{+++}$ (see table 9 and A6 of ref.\cite{west05}),
one can show that in  these configurations Eq.(\ref{condb}) and Eq.(\ref{condc}) are not
satisfied. We have thus again a perfect agreement between the existence of
intersecting extremal brane solutions in the maximally oxidised $F_4$ theory and the
existence of the algebraic counterpart in $F_4^{+++}$.  This
concludes the proof of our central theorem 3.

\section{Extension to the exotic phases}

In superstring theories, the $U$-duality group corresponds to the Weyl group of
$E_8^{+++}$
\cite{eli,obe,ban,ehtw}. In the type $IIA$ language the non-trivial Weyl reflection
generated by the simple root which do not belong to the gravity line of $E_8^{+++}$
corresponds to a double $T$-duality in the $\hat 9$ and $\widehat{10}$ directions plus an
exchange of these two directions. Combining this Weyl reflection with the ones of the
gravity line one is inevitably led to consider
$T$-duality involving the timelike direction. Compactification of the timelike direction in
string theories \cite{compt} along with timelike $T$-dualities
\cite{exop} have been considered. It has been shown that dualities involving the timelike
direction can change the signature of space-time \cite{exop} and lead to `exotic' phases
of M-theory with  more than one time. Starting with the `orthodox' M-theory
corresponding to 11-dimensional supergravity with signature
$(T,S)=(1,10)$ it has been shown \cite{exop} that  one can reach by $U$-duality
$M^*$-theory   with $(T,S)=(2,9)$ and the `wrong' sign in front of the kinetic
term of the four-form  field strength $F_4$. One can  also reach $M^\prime$-theory
with 
$(T,S)=(5,6)$  and the conventional sign in front of the kinetic term of $F_4$. In the
$E_8^{+++}$ theory (and more generally in $\G$) the usual signature of space-time  is
implemented through the temporal involution \cite{eh} which ensures that in the $A_{10}$
($A_{D-1}$) level  decomposition  we have
$SO(1,10)$ ($SO(1,D-1)$) tensors. The expected existence of the exotic phases in the
$E_8^{+++}$ context has been shown in ref.\cite{keu}.  The author
studied the Weyl  reflections and showed that starting with the temporal involution one
can reached the above-mentioned exotic phases using Weyl reflections involving the time
direction.  In other words, in contrast with  the usual Cartan involution, the
temporal involution is not invariant under conjugation by all the Weyl reflections. As a
consequence, the $E_8^{+++}$-invariant theory proposed in \cite{eh}
contains also  the algebraic  counterpart of single extremal brane solutions of the exotic
phases $M^*$ and $M^\prime$.

The argument can be generalised to all $\G$ theories. Indeed, a subset of the Weyl
reflections of $\G$  maps extremal branes onto other extremal branes generalising to all
`M-theories' the notion of dualities \cite{eh}.  As pointed out in \cite{keu}, we can expect
that some Weyl reflections  (around roots not on the gravity line and involving time
direction) will not leave the temporal involution invariant and will lead to `exotic' $\G$
theories. These would  correspond to maximally oxidised $\cal G$ with some $(T,S)$
signatures and with possibly wrong sign  kinetics terms for some form field strengths.

Even without having classified explicitly\footnote{For very-recent results on the
exotic phases of  the $\cal G$ theories see \cite{keu2} appendix A.} the possible exotic 
phases for all $\G$, we want to argue here that  when such phases do exist the analysis of 
the previous section extend to them. We can first study the extremal branes of these
maximally oxidised $\cal G$ theory with  some fixed signature $(T,S)$. Extremal branes of
the exotic phases of M-theory ($E_8$) have been considered in
\cite{exob}. The existence of intersecting brane solutions for a generic  theory in
$D=T+S$ dimensions with $T$ timelike dimensions and $S$ spacelike dimensions  which
includes gravity, a dilaton, form field strengths $F_{n_I}$ of arbitrary degree $n_I$ with
arbitrary couplings to the dilaton $a_I$ and  `$n_I$-form'  kinetic terms with an arbitrary
sign given by $\Theta_I=\pm 1$ ($\Theta = +1$ corresponding to the conventional sign)
has been studied in ref.\cite{ah}. Each single extremal $q_A$-brane  is characterised by
$s_A$ spatial longitudinal directions and $t_A$ temporal longitudinal directions. There is 
furthermore a  condition which has to be satisfied in order for the single brane solution 
to exist \cite{exob}:
\begin{equation}
\Theta_A (-1)^{t_A+1}=1\, ,
\label{theta}
\end{equation} where $\Theta_A =\Theta_{I=q_A+2}$ associated with the $F_I$ in the
action when the $q_A$-brane  is an electric one and 
$\Theta_A=(-1)^{T+1} \Theta_{I=D-q_A-2}$ when the $q_A$-brane is a magnetic one. The
condition Eq,(\ref{theta}) is trivially satisfied in the orthodox phases. The generalised
intersection rules give then for each pair $(A,B)$ of extremal branes the following 
conditions involving  the number of common spacelike directions  $\bar s$  and the
number of common  timelike directions $\bar t$ \cite{ah}
\begin{equation}
\bar{s}+\bar{t}=\frac{(s_A+t_A)(s_B+t_B)}{D-2}-\frac{1}{2}
\varepsilon_A a_A \varepsilon_B a_B \label{exorule}\, .
\end{equation} We note that the generalised intersection rule Eq.(\ref{exorule}) depends
only on the total dimensions
$s_A+t_A$, $s_B+ t_B$ and $\bar{s} +\bar{t}$ irrespectively of the temporal or spatial
nature of them. It does not depend on the signature nor on the sign of the kinetic terms
of the forms.  Consequently in $\G$ language Eq.(\ref{exorule}) can still be translated
into the orthogonality condition proved in theorem 1,  independent of the involution.  Only
the existence of a solution for the building blocks, namely the single extremal branes, 
depends on  the particular involution through Eq.(\ref{theta}).  Each   single extremal
$q_A$-brane  is characterised in 
$\G$ by only one non-zero field component $A_{\tau_1 \dots \tau_{t_A}\lambda_1 \dots
\lambda_{s_A}}$  where $\tau_i$ are timelike directions and $\lambda_i$ spacelike ones. 
This field is the parameter of a  antisymmetric tensor  step operator of low level 
$R^{\tau_1 \dots \tau_{t_A} \lambda_1 \dots
\lambda_{s_A}}$ and  corresponds to the  real positive root
$\alpha_{(\tau_1 \dots \tau_{t_A},\lambda_1, \dots ,\lambda_{s_A})}$. 
All the analysis of section 2
can thus  be repeated in this framework.  Under a Weyl reflection transforming
 an orthodox intersecting brane configuration into an exotic one, the
invariance of the lapse constraint Eq.(\ref{xiextremal}) , and particularly the sign of the
last term is ensured by the condition  Eq.(\ref{theta}).

Thus to each  intersecting  extremal brane 
configuration of an `exotic' maximally oxidised $\cal G$ theory   there exists an algebraic
counterpart which is an exact solution of the $\G$ theory.

\section*{Acknowledgments}

This work was supported in part  by the NATO grant PST.CLG.979008,
   by the ``Actions de Recherche Concert\'ees'' of the ``Direction de la   Recherche
Scientifique - Communaut\'e Fran\c caise de Belgique, by a ``P\^ole   d'Attraction
Interuniversitaire'' (Belgium), by IISN-Belgium (convention 4.4505.86),   by Proyectos
FONDECYT 1020629, 1020832 and 7020832 (Chile) and by the   European Commission RTN
programme HPRN-CT00131, in which F.~E and L.~H. are associated to the Katholieke
Universiteit te Leuven (Belgium).

\end{document}